\documentclass[aps,prl,groupedaddress, twocolumn,  nofootinbib]{revtex4-1}
\usepackage{graphicx, amsmath, bbm, hyperref}

\begin{document}

\title{Numerical evidence for a phase transition in 4d spin foam quantum gravity}

\author{Benjamin Bahr}
\email[]{benjamin.bahr@desy.de}

\author{Sebastian Steinhaus}
\email[]{sebastian.steinhaus@desy.de}
\affiliation{II. Institute for Theoretical Physics\\University of Hamburg\\ Luruper Chaussee 149\\22761 Hamburg\\Germany}

\date{\today}

\begin{abstract}

Building on recent advances in defining Wilsonian RG flows, and in particular the notion of scales, for background-independent theories, we present a first investigation of the renormalization of the $4d$ spin foam path integral for quantum gravity, both analytically and numerically. Focussing on a specific truncation of the model using a hypercubic lattice, we compute the RG flow and find strong indications for a phase transition, as well as an interesting interplay between the different observed phases and the (broken) diffeomorphism symmetry of the model. Most notably, it appears that the critical point between the phases, which is a fixed point of the RG flow, is precisely where broken diffeomorphism symmetry is restored, which suggests that it might allow for the definition a continuum limit of the quantum gravity theory.
\end{abstract}


\maketitle

\section{Motivation}

The unification of the theory of general relativity with the principles of quantum physics is one of the great challenges for theoretical physics. Although there are several candidates for such a theory of quantum gravity, none of them is feature complete.

One of the further developed ones is represented by the  so-called spin foam models (SFM). Based on a path-integral formulation of loop quantum gravity \cite{Reisenberger:1996pu}, SFM have seen major advances in recent years \cite{Perez:2012wv, LectureNotesCarlo2013}. However, one of the crucial open questions of SFM is that of renormalization, in particular in relation to their continuum limit \cite{Rovelli:2010qx}, and the well-known non-renormalizability of perturbative quantum gravity \cite{Goroff:1985sz}.

The reason that the renormalization of SFM is still open is connected to the fact that renormalization describes the behaviour of a theory at different scales, encoded in the running of its coupling constants \cite{Wilson:1973jj}. In SFM, these questions are hard to tackle, because of the background-independent nature of the theory: since any geometric information about scales, such as latticle lengths, are encoded in the field itself, standard renormalization techniques are not directly applicable.


However, recently work in the area of background-independent renormalization has seen significant advances. Using tools from tensor networks \cite{guwen,levin}, adapted to the setting in quantum gravity, has allowed to formulate the RG flow also for spin foam models \cite{Bahr:2012qj, Dittrich:2013voa, Bahr:2014qza}. These methods have been applied to analogue models \cite{Dittrich:2013bza, Dittrich:2014mxa}, revealing a rich phase structure. \footnote{Complementary work on renormalizing the theory have been carried out in \cite{Riello:2013bzw, Banburski:2014cwa}.}

In this article, we present a first application of these techniques to the SFM for 4d, Euclidean quantum gravity \cite{Engle:2007wy, Freidel:2007py}. Although we are working a in a certain truncation of the model, we find a nontrivial renormalization group flow, as well as a critical fixed point, with intriguing physical and geometrical properties.

\section{Spin foam quantum gravity on a hypercubic lattice}

%
%
%

The SFM is a path integral, defined on a discretization of space-time, in particular on the dual of a cellular decomposition of the space-time manifold which we choose to consist of 4d hypercuboids.\footnote{The original model by Engle, Pereira, Rovelli and Livine \cite{Engle:2007wy} and Freidel and Krasnov \cite{Freidel:2007py} was defined for a simplicial decomposition. We use the generalization by Kaminski, Kisielowski and Lewandowski \cite{Kaminski:2009fm}, which works for arbitrary polyhedral decompositions of space-time.} We emphasize that this lattice is purely combinatorial, and does not contain intrinsic geometric information (such as e.g.~a lattice spacing).

A quantum geometry on this lattice is defined by a ``state'', i.e.~an assignment of half-integers (``spins'') $j_f$ to the 2d ``squares'' $f$, and invariant tensors (``intertwiners'') $\iota_e$ to the 3d ``cuboids'' $e$ of the 4d lattice. The spin $j_f$ corresponds to the area of the square $f$, while the intertwiner $\iota_e$ determines the 3d shape of the cuboid. The weight of such a state is given by amplitudes $\mathcal{A}_f$, $\mathcal{A}_e$, and $\mathcal{A}_v$ for squares $f$, cuboids $e$ and hypercuboids $v$.\footnote{In the SFM literature, these are called face-, edge- and vertex amplitudes. This nomenclature arises from the dual lattice, where faces are dual to 2d squares, edges are dual to 3d cuboids, and vertices are dual to 4d hypercuboids.} These functions depend locally on the state. The path integral is then given by a sum over all possible states $(j_f,\iota_e)$, i.e.

\begin{eqnarray}\label{Eq:StateSum}
Z=\sum_{j_f,\iota_e}\prod_f\mathcal{A}_f\prod_e\mathcal{A}_e\prod _v{\mathcal{A}}_v\;.
\end{eqnarray}

\noindent It is quite an ambitious endeavour to consider the full sum (\ref{Eq:StateSum}). In all practical situations, it will be prudent to restrict it to only a few, dominant states, in order to derive predictions from the path integral. This is common practice in lattice gauge theory, while for SFM, a good intuition about which states $(j_f,\iota_e)$ dominate the sum (\ref{Eq:StateSum}) has not yet been fully developed. In \cite{Bahr:2015gxa}, a subset of states in question has been introduced, which should contribute significantly to the Minkowski vacuum of the theory.
\begin{figure}
\includegraphics[width=0.3\textwidth]{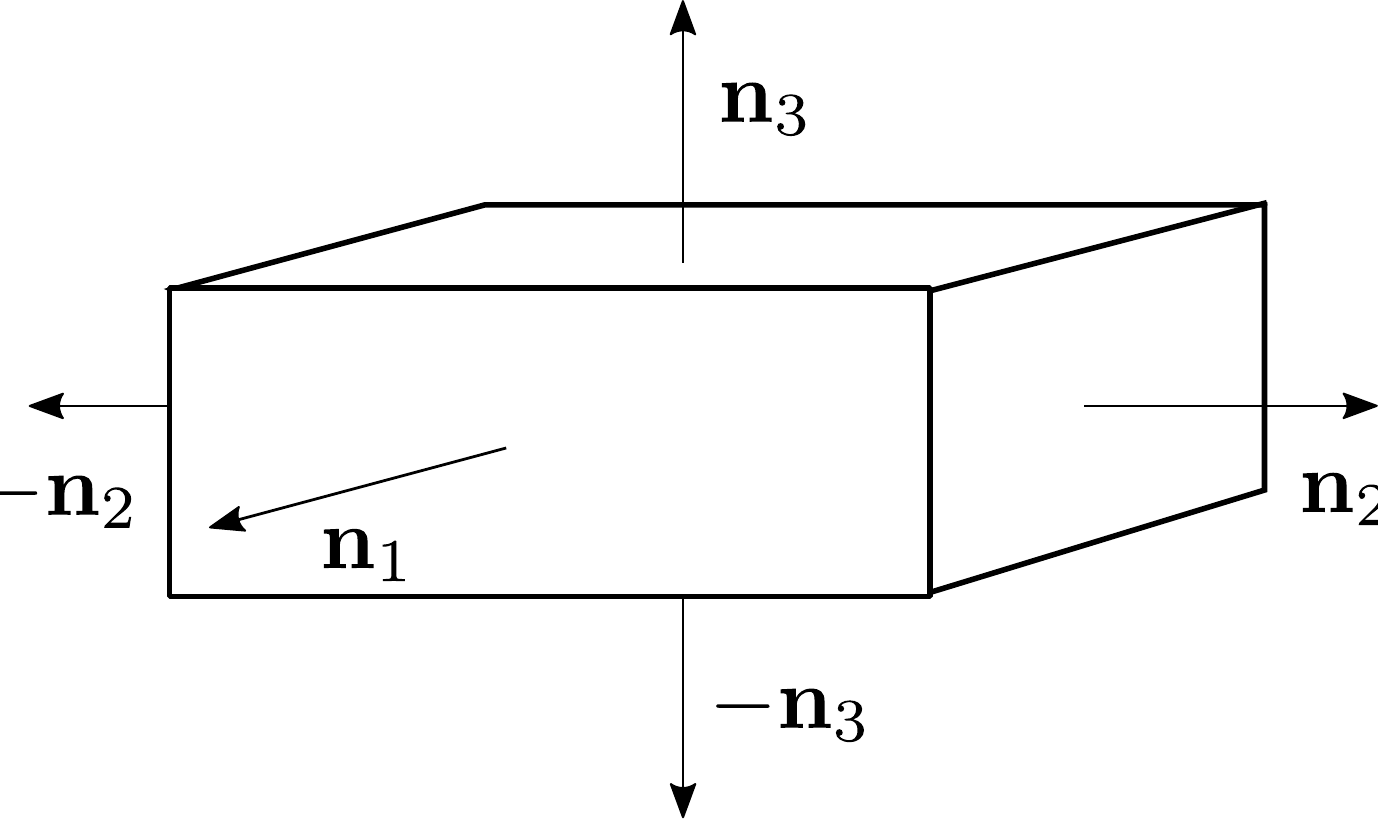}
\caption{Visualization of coherent intertwiner $\left | \iota_{j_1,j_2,j_3} \right \rangle$.
\label{fig:cuboid}
}
\end{figure}
Their geometric interpretation is adapted to the 4d hypercubic lattice, taking its symmetries into account. The intertwiners $\iota_e$ are restricted to be \emph{quantum cuboids}, i.e.~they are quantum versions of classical 3d cuboids (see figure \ref{fig:cuboid}). The spins are restricted to be such that they agree with the cuboidal symmetry. This still allows for local degrees of freedom, capturing non-topological effects of the whole path integral, while at the same time making the sum much more manageable.

The quantum cuboid intertwiner states depend on three spins, as a cuboid is completely determined by its three areas. They are given as group averages over six Livine-Speziale coherent states $|j\,{\rm \bf n}\rangle=g({\rm \bf n})|j\,j\rangle$, where $g({\rm \bf n})$ is an $SU(2)$ element rotating the $z$-axis towards the direction of the vector ${\rm \bf n}$ \cite{Livine:2007vk, Livine:2007ya}:

\begin{eqnarray}\label{Eq:CoherentCuboidIntertwiner}
\left | \iota_{j_1,j_2,j_3} \right \rangle = \int_{\text{SU}(2)} dg\;g\triangleright\bigotimes_{i = 1}^3 \left | j_i \, {\rm\bf n}_i \right \rangle \otimes \left | j_i\, -{\rm\bf n}_i \right \rangle,
\end{eqnarray}
\noindent where ${\rm \bf n}_i$ form an orthonormal system in $\mathbbm{R}^3$.

The combinatorics of the lattice imply that a piece of 4-dimensional space-time (i.e.~a 4d hypercuboid in the lattice) has a three-dimensional boundary consisting of eight cuboids of the type (\ref{Eq:CoherentCuboidIntertwiner}). The path integral amplitude $\mathcal{A}_v$ for this hypercuboid is given by the contraction of boosted intertwiners \cite{Engle:2008ev}. The edge amplitudes $\mathcal{A}_e$ are taken to be the inverse norm squared of the coherent intertwiners, i.e. $\mathcal{A}_e=\|\iota_{j_1,j_2,j_3}\|^{-2}$, and the face amplitudes $\mathcal{A}_f$ are chosen to be
\begin{eqnarray}\label{Eq:FaceAmplitude}
\mathcal{A}_f\;=\;\Big((2j_f^++1)(2j_f^-+1)\Big)^\alpha      
\end{eqnarray}

\noindent where $j_f^\pm=\frac{|1\pm \gamma|}{2}j_f$, and $\gamma$ is the Barbero-Immirzi parameter. We have also introduced $\alpha$ as a free parameter. The dependence of the model on the dimensionless coupling constant $\alpha$ has been investigated in \cite{Bahr:2015gxa}. The analysis in this article will reveal a renormalization group flow in $\alpha$.

Due to the combinatorial symmetry of the lattice, the face- and edge amplitudes $\mathcal{A}_e$, $\mathcal{A}_f$ can be absorbed into the vertex amplitudes $\mathcal{A}_v$, allowing to rewrite (\ref{Eq:StateSum}) as
\begin{eqnarray}\label{Eq:TruncatedStateSum}
Z=\sum_{j_f}\prod _v\hat{\mathcal{A}}_v^{(\alpha)}
\end{eqnarray}

\noindent In the restricted state sum (\ref{Eq:TruncatedStateSum}), the intertwiners $\iota_e$ are completely determined by the spins. Since we will be concerned with the sum over spins along a vast range, it is a reasonable approximation to replace the amplitudes by their large-$j_f$ asymptotic expression, turning sums into integrals.\footnote{We employ methods by Barrett et al \cite{Barrett:2009gg}, and Freidel and Conrady \cite{Conrady:2008mk}.} In this limit, the dependence of both Newton's constant $\kappa$ and the Barbero-Immirzi parameter $\gamma$ vanishes for quantum cuboids \cite{Bahr:2015gxa}, which is why $\alpha$ is the only coupling constant remaining.

Note that the dressed amplitude $\hat{\mathcal{A}}_v^{(\alpha)}$ depends on 6 boundary spins (areas of squares), rather than 4 edge lengths. The excess degrees of freedom correspond to ``twisted'' geometries, and are discussed in \cite{Freidel:2010aq, Freidel:2013bfa}.

%

\section{Coarse graining hypercuboids}

We consider a nuclear step of the RG flow of the model (\ref{Eq:TruncatedStateSum}), similar to a block spin transformation. This is the coarse graining step from $2\times 2\times 2\times 2=16$ fine 4d hypercuboids $v_i$ to a coarse one $V$. The crucial part of this analysis is the relation of the boundary data of the 16 fine hypercuboids (24 spins) to the boundary data of the coarse hypercuboid (6 spins).

This relation is given by the so-called embedding map $\Phi_{b'b}$, which maps states on the coarse boundary $b$ to states on the fine boundary $b'$. The precise choice for $\Phi_{b'b}$ is quite nontrivial, and greatly influences the details of the RG flow. Understanding the role the maps $\Phi_{b'b}$ play, prominently by Dittrich and others \cite{Dittrich:2012jq, Dittrich:2013bza, Dittrich:2013xwa, Dittrich:2014ala, Livine:2013gna}, has been one of the corner stones of advances in this area.

For our analysis, we choose embedding maps $\Phi_{b'b}$ adapted to the geometrical interpretation of the quantum cuboid boundary states (\ref{Eq:CoherentCuboidIntertwiner}). Since, in the semiclassical regime, the spins $j_f$ are proportional to areas of squares $f$, it seems reasonable to demand that four fine spins should add up to one coarse one. This way, an intertwiner on the coarse boundary, with spin $J_F$, can be identified with a tensor product of the highest weight vector of the tensor product of four fine spins $j_{f}$, where the four fine squares $f$ comprise the coarse face $F$. In other words,
\begin{equation} \label{Eq:CoarseGrainingSpins}
C_F\;:=\; J_F - \sum_{f\subset F} j_{f} \;\stackrel{!}{=}\;0\; .
\end{equation}

\begin{figure}[hbt!]
\includegraphics[width=0.43 \textwidth]{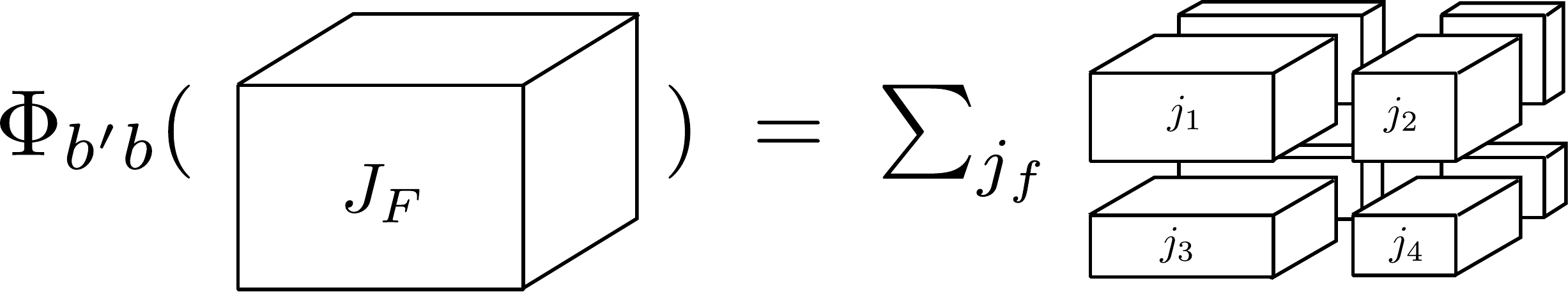}
\caption{Graphical representation of the embedding maps. The sum ranges over all $j_f$ so that e.g.~$j_1+j_2+j_3+j_4=J_F$.\label{Fig:EmbeddingMap}}
\end{figure}

\noindent Clearly there are many different values $j_f$ which satisfy (\ref{Eq:CoarseGrainingSpins}) for a given $J_F$. The embedding map $\Phi_{b'b}$ should therefore map to a coherent superposition of all of these possibilities, which we have depicted schematically in figure \ref{Fig:EmbeddingMap}.

With this embedding map and an amplitude $\hat{\mathcal{A}}_v^{(\alpha)}$, the renormalized amplitude $\mathcal{A}_V '$, regarded as linear form on the boundary Hilbert space, is just $\mathcal{A}_V'\;=\;(\otimes_{i=1}^{16}\hat{\mathcal{A}}_{v_i}^{(\alpha)})\Phi_{b'b}$. In terms of the amplitude functions, it is given by the integral over the fine spins $j_f$:

\begin{eqnarray}\label{Eq:RenormalizationStep}
\mathcal{A}'_{V}(J_F) \;&=&\; \int dj_f \,N_{J_F}^{j_f}\,\prod_F\delta(C_F) \prod_{i=1}^{16}\hat{\mathcal{A}}^{(\alpha)}(j_f) \nonumber\\[5pt]
\end{eqnarray}

\noindent where the $N_{J_F}^{j_f}$ are numerical coefficients which take care of over-/undercounting of states due to the symmetry restriction.

\subsection{Truncation of the RG flow}

\noindent Usually in an RG step many more couplings are generated, so the renormalized amplitude $\hat{\mathcal{A}}_V'$ will not necessarily be a quantum hypercuboid again. Geometrically, the path integral does not generate curvature, so a flat $4d$ hypercuboid should still be a good approximation of the renormalized vertex $V$. Therefore, we truncate the flow to the space of hypercuboids, expecting a not too aggravating error in that approximation.

Technically the truncation process is performed as described in  \cite{Bahr:2014qza}, by comparison of observables. The renormalized amplitude (\ref{Eq:RenormalizationStep}) is thereby replaced by a hypercuboidal amplitude $\hat{\mathcal{A}}_V^{(\alpha')}$, which depends on the coarse spins $J_F$. The ``small scale'' fluctuations of the fine spins $j_f$ are absorbed in the change of the coupling constant $\alpha\to\alpha'$. The value of $\alpha'$ is then such that the expectation values
\begin{eqnarray}\label{Eq:ObservableExpectationValue}
\langle\mathcal{O}\rangle\;=\;\frac{1}{Z}\sum_{j_f}\; \mathcal{O}(j_f)\;\prod_{v}\hat{\mathcal{A}}_v^{(\alpha)}
\end{eqnarray}

\noindent of certain previously chosen obervables $\mathcal{O}$ are as similar as possible on the coarse lattice, compared to taking $\alpha$ on the fine lattice. This determines the renormalized value of $\alpha'$, depending on $\alpha$. Iterating these steps then generates the flow of the coupling constant.

\subsection{The renormalization step}

Since we only consider one coupling constant, comparing one observable $\mathcal{O}$ is sufficient to define the flow. For this we consider a lattice of two hypercuboids $V_1$ and $V_2$, with fixed total volume ${\rm Vol}_1+{\rm Vol}_2=1$. The observable we choose is the variance of the volume of one of the hypercuboids
\begin{eqnarray}\label{Eq:VolumeVariance}
\mathcal{O}\;:=\;\Delta {\rm Vol}_1\;=\;\big({\rm Vol}_1\,-\,\langle {\rm Vol}_1\rangle\big)^2\;.
\end{eqnarray}

\noindent Symmetry dictates that $\langle {\rm Vol}_1\rangle=\frac{1}{2}$ for all $\alpha$. However, a posteriori one finds that the variance (\ref{Eq:VolumeVariance}) is an excellent ordering parameter for $\alpha$, since it behaves monotonously as the coupling constant varies \cite{Bahr:2015gxa}.

The actual integrals are being carried out using standard Monte Carlo methods, which are especially adapted to approximate high-dimensional integrals.

\begin{figure}
\includegraphics[width= 0.43 \textwidth]{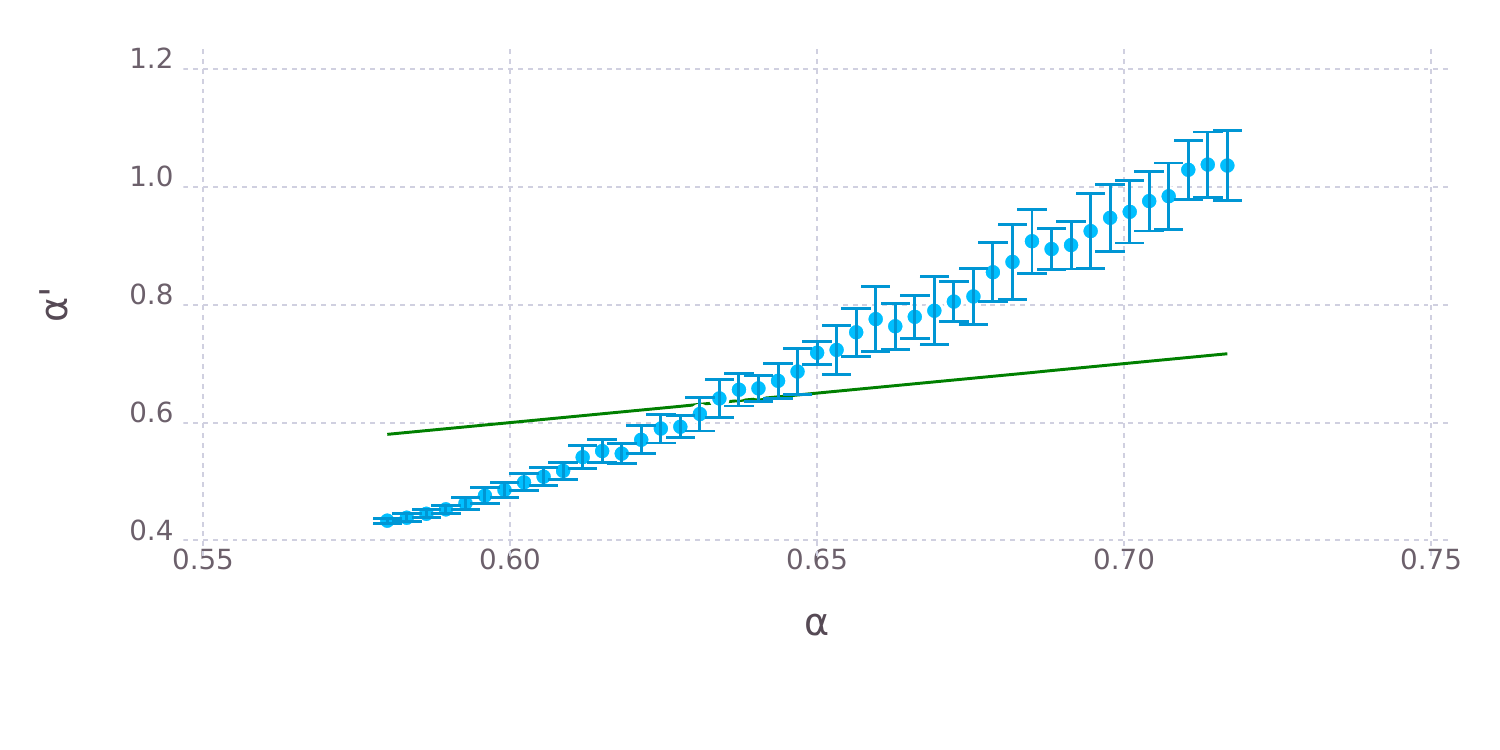}
\caption{Renormalized $\alpha'$ depending on $\alpha$. The straight line is $\alpha'=\alpha$. Note the unstable fixed point around $\alpha \approx 0.64$, where the two curves intersect. Error bars come from the finite precision of Monte Carlo methods.
\label{Fig:RGFlow}}
\end{figure}

The result of the computation of a renormalization group step for various different values of $\alpha$ has been depicted in figure \ref{Fig:RGFlow}.

It is apparent that the flow in $\alpha$ has a fixed point, which we find numerically to be
\begin{eqnarray}
\alpha_c\;\approx\;0.64\,\pm\,0.04\,,
\end{eqnarray}
\noindent which sits at the intersection of the computed curve $\alpha'(\alpha)$ with the line $\alpha'=\alpha$. Since the slope of $\alpha'(\alpha_c)>1$, this fixed point is unstable, i.e.~UV attractive.

The existence of this fixed point is remarkable, in particular because it occurs despite our radical truncation of the model. In the following, we analyse its geometric properties, and elucidate its connection to the diffeomorphism on the lattice.

\section{Properties of the fixed point}

\subsection{Dominant states in the path integral}

In order to understand the presence of the fixed point $\alpha_c$ in the RG flow, we take a closer look at the integral in (\ref{Eq:RenormalizationStep}). In particular, it is interesting to consider the dominant fine configurations $\{j_f\}_f$ of (\ref{Eq:RenormalizationStep}), depending on $\alpha$. To this end, we employ a tactics which is also widely applied in Causal Dynamical Triangulations \cite{Ambjorn:2004qm, Ambjorn:2011cg, Ambjorn:2013apa}. We consider a random walker through state space, which has a bias towards more probable configurations. If one lets the walker run for long enough, it will settle for a local maximum, and fluctuate around it. This maximum will be one of the dominant configurations of the path integral, and we will interpret its geometry.

During the course of this random walk, we keep track of one fine spin $j_i$, while we keep all coarse boundary spins $J_F$ fixed.

\begin{figure}
\includegraphics[width=0.43 \textwidth]{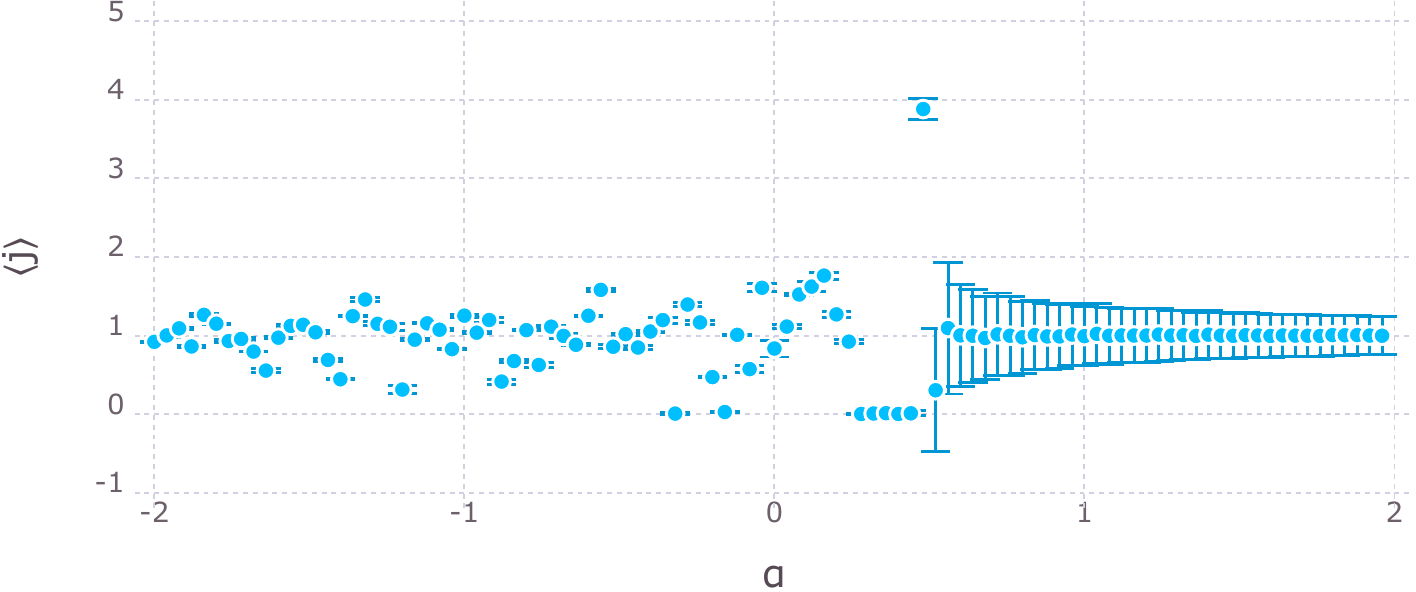}
\caption{Average value of fine spin $j_i$ over a biased random walk with $10^6 $ steps, as a function of $\alpha$. Error bars indicate the fluctuations of $j_i$. All coarse $J_F = 4$.
\label{Fig:RandomWalker}}
\end{figure}

In figure \ref{Fig:RandomWalker} we show the average value of the fine spin $j_i$ during the random walk. One clearly recognizes qualitatively different behaviour of the walker, depending on the value of $\alpha$.

The interpretation is as follows: For larger values of $\alpha$, the integrand in (\ref{Eq:RenormalizationStep}), i.e.~the product of 16 hypercuboidal amplitudes $\hat{\mathcal{A}}_{v_i}^{(\alpha)}$, is peaked around the configuration $j_f=\frac{J_F}{4}$. This corresponds to the geometry where the coarse hypercuboid is subdivided into 16 finer hypercuboids \emph{in a regular way}. The width of that peak gets larger, as $\alpha$ approaches a critical value from above, which we find here to be around $\alpha_c\approx 0.56$.

For smaller values of $\alpha$, the random walker registers values of $j_i$ in a broad range between $0$ and $J_F$, with small individual fluctuations. This is an indicator that the maximum of the amplitude does not lie at the regular subdivision, but that there are several ``off-center'' peaks. Since $\langle j_i\rangle_{\rm exact} = J_F/4$, these peaks have to be arranged symmetrically around the regular case. But due to the sharpness of the peaks (signified by the small fluctuations), the random walker is unlikely to escape, as soon as it has settled for one.

The geometry of these peaks has to be interpreted as subdivisions of the $4d$ hypercuboid $V$ which are \emph{irregular}, since $j_i\neq \frac{J_F}{4}$ at the dominant configuration. It should be noted that an irregular subdivision is diffeomorphically equivalent to a regular one, however.

So the main contribution to the path integral (\ref{Eq:RenormalizationStep}) comes from quite different (albeit diffeomorphically equivalent) states $\{j_f\}_f$, depending crucially on the value of $\alpha$.

Note that exactly the same behaviour described here has been found in \cite{Bahr:2015gxa}, but for 2 hypercuboids instead of 16. The behaviour is intricately connected to the (non-) breaking of diffeomorphism symmetry in the path integral.

\subsection{Diffeomorhism symmetry in the path integral}

The group of space-time diffeomorphisms is a gauge-symmetry of general relativity, incorporating the principle of general covariance. In discrete gravity approaches like Regge Calculus it is usually replaced by the ``vertex-displacement symmetry'', which captures the analogous gauge degrees of freedom in the discrete setting \cite{FreidelLouapreDiffeo2002, Dittrich:2008pw}. This symmetry is, however, broken in the 4d version of the theory \cite{Bahr:2009ku, Bahr:2009qc}. Since the large $j$-limit of the EPRL-FK spin foam model is dominated by the Regge action \cite{Barrett:2009gg, Conrady:2008mk, Magliaro:2011dz}, one would expect the same  breaking to arise in spin foam quantum gravity in some form. This has been demonstrated for the large $j$-limit of the symmetry-restricted spin foam state sum \cite{Bahr:2015gxa}, and we find its effect again here in the analysis of the RG flow.

Consider a 4d hypercuboid with fixed boundary, subdivided into two hypercuboids $v_1$ and $v_2$. All possible subdivisions are diffeomorphically equivalent, so the issue is connected to the behaviour of their total amplitude
\begin{eqnarray}\label{Eq:TwoVertexAmplitudes}
I(\alpha,x)\;=\;\hat{\mathcal{A}}_{v_1}\hat{\mathcal{A}}_{v_2}\;,
\end{eqnarray}

\begin{figure}
\includegraphics[width=0.43 \textwidth]{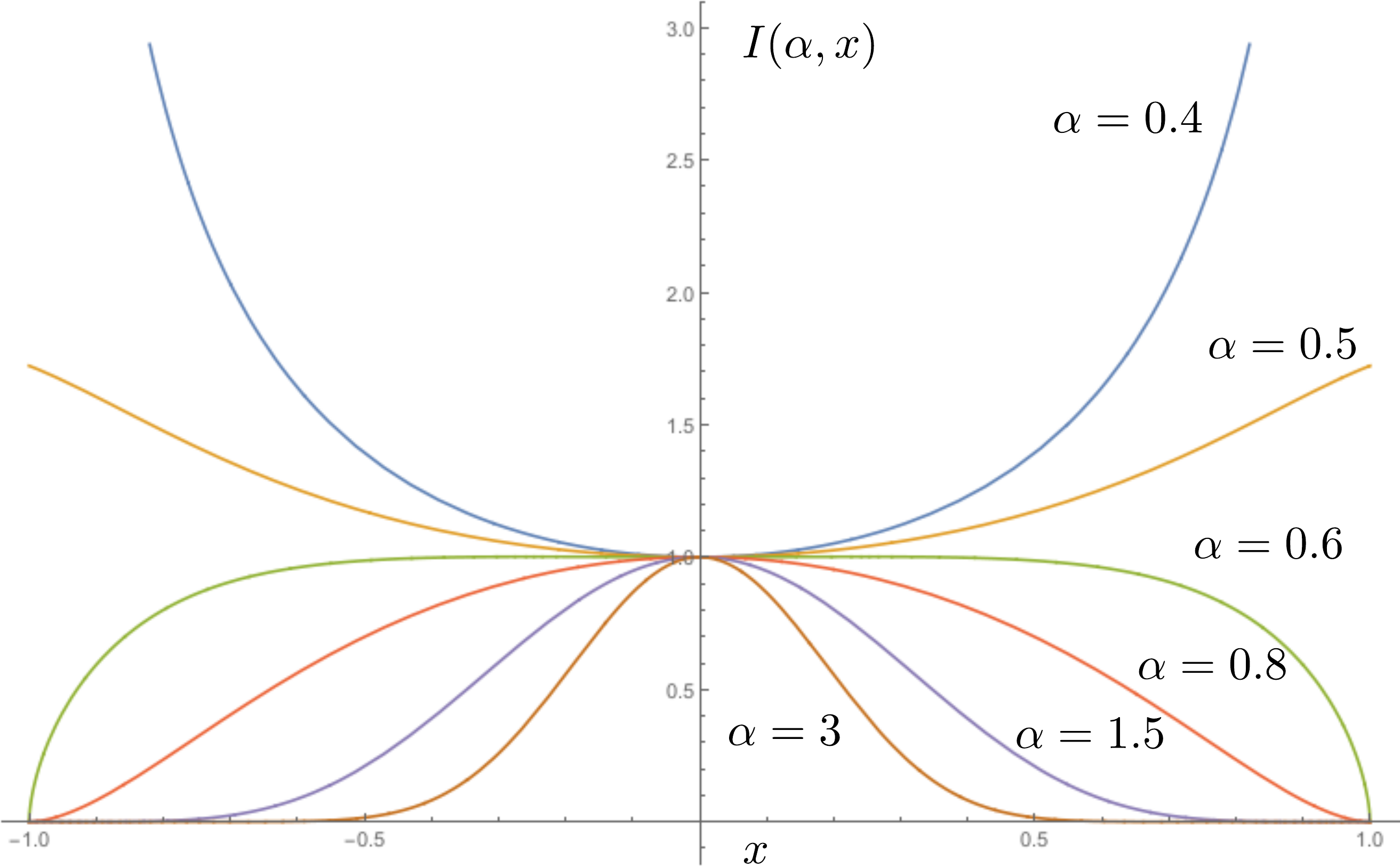}
\caption{Product (\ref{Eq:TwoVertexAmplitudes}) of two vertex amplitudes, depending on the subdivision of one $4d$ hypercuboid into two. The parameter $x$ captures the irregularity of the subdivision. Curves are shown for different values of $\alpha$.
\label{Fig:DiffeoSymmetryPeaks}}
\end{figure}

\noindent where $x\sim\frac{{\rm Vol}_1-{\rm Vol_2}}{{\rm Vol}_1+{\rm Vol_2}}$ parameterizes the deviation from the regular subdivision. Vertex translation manifests itself by varying $x$. In \cite{Bahr:2015gxa} it has been found that the $I(\alpha, x)$ is invariant under vertex translation symmetry if and only if the coupling constant $\alpha$ attains the critical value
\begin{eqnarray}
\alpha_c\;=\;\frac{91}{150}\;\approx \;0.61\;,
\end{eqnarray}

\noindent which can be seen in figure \ref{Fig:DiffeoSymmetryPeaks}. The figure also shows that for $\alpha<\alpha_c$, the dominant state is those where the subdivision is highly irregular (i.e.~$x\approx\pm 1$), while for $\alpha>\alpha_c$, it becomes more and more centered around the regular subdivision at $x=0$.

Figure \ref{Fig:RandomWalker} shows that the behaviour persists for 16 hypercuboids, instead of 2.

\subsection{Geometric interpretation}

This elucidates the geometric interpretation of the fixed point $\alpha_c$: Disregarding non-geometric configurations (which are suppressed in the path integral \cite{Bahr:2015gxa}), the integral (\ref{Eq:RenormalizationStep}) ranges over all possible decompositions of the coarse $4d$ hypercuboid with fixed geometry into 16 fine ones. The value of $\alpha$ influences the path integral measure, which generically does not treat all these decompositions equally, i.e.~breaks diffeomorphism symmetry. For $\alpha>\alpha_c$ the dominant configuration is the regular subdivision, while for $\alpha<\alpha_c$ it appears to be a linear superposition of irregular subdivisions, arranged symmetrically around the regular one.

As has been shown in \cite{Bahr:2015gxa}, for $\alpha\to\alpha_c$, the width of the peaks around the dominant contributions grows, so that at the critical point all configurations contribute (approximately) the same, i.e.~diffeomorphism symmetry is restored.


\section{Discussion and Outlook}

In this article we have detailed a renormalization procedure in spin foam quantum gravity. We have worked in a symmetry-restricted truncation of the Riemannian signature EPRL-FK model, with one coupling constant $\alpha$. Though this analysis incorporated only a subset of the degrees of freedom of 4d quantum gravity, the analysis already showed very promising results:


\begin{itemize}
\item The flow in $\alpha$ has a fixed point at $\alpha_c\approx 0.64\pm 0.04$, to our numerical precision. Computing the RG step with values of $\alpha$ in this region confirms the fixed point properties of the flow.
\item This fixed point is UV-attractive. It shares this property with the non-Gaussian fixed points encountered in the Asymptotic Safety Scenario \cite{Niedermaier:2006wt}.
\item It coincides, within our numerical precision, with the critical point observed in \cite{Bahr:2015gxa}. We could demonstrate that $\alpha_c$ separates two regions with qualitatively very different behaviour of the amplitude function $\hat{\mathcal{A}}_v^{(\alpha)}$. It is apparent that this difference influences the characteristics of the flow dramatically.
\item Very importantly, the critical point plays a crucial role for the vertex displacement symmetry (figure \ref{Fig:DiffeoSymmetryPeaks}), since it is the one and only value of the coupling constant $\alpha$, at which this symmetry is, to a very good extent, present. To us this appears to be the main reason for the presence of this fixed point.
\end{itemize}

These findings are exciting and very encouraging for us, since they foster hope that, as has been conjectured for some time \cite{Dittrich:2008pw}, diffeomorphism symmetry of the quantum gravity path integral might be restored at the renormalization group fixed point (see also \cite{Bahr:2009ku, Bahr:2009qc, Bahr:2011xs} for some discussion).

Now it needs to be tested whether these features hold in general, or whether they are an artifact of the approximations we have used. The next steps will therefore be to remove some of the restrictions we subjugated our analysis to, and to repeat the analysis of this article. This means including more degrees of freedom, more coupling constants, as well as working beyond the large $j$-regime of state space.

If one were to find that the fixed point persists, it could provide a way to define the continuum limit of spin foam models, defining a possibly renormalizable quantum gravity theory in the continuum.

\vspace{1. cm}

\begin{acknowledgments}
{\bf Acknowledgments:} This work was funded by the project BA 4966/1-1 of the German Research Foundation (DFG). S.S.~would like to thank Erik Schnetter for introducing him to the programming language julia, with which the numerical computations were carried out.
\end{acknowledgments}

\bibliography{bibliography}

\end{document}